\documentclass[aps,prl,preprint,groupedaddress]{revtex4-2}

\usepackage[utf8]{inputenc}
\usepackage{graphicx}
\usepackage{subfig}
\usepackage{caption}
\usepackage{float}
\begin{document}


\title{Predictive design of two-dimensional electrides with tunable magnetic, topological, and superconducting properties}


\author{Haomin Fei, Ping Cui and Zhenyu Zhang}
\affiliation{International Center for Quantum Design of Functional Materials (ICQD), University of Science and Technology of China, Hefei, Anhui 230026, China}


\date{\today}

\begin{abstract}
Two-dimensional materials are of interest for their exotic properties, for example, superconductivity, and highly tunability. Focusing on phonon-mediating superconductivity, one would propose to promote critical temperature by substituting heavy elements by lighter ones, in order to increase Debye temperature. Following recent experimental progress in transition-metal nitrides and theoretically revealing W$_2$N$_3$ as a candidate for high-temperature superconductivity, we investigate the possibility of two-dimensional superconductivity through surface engineering on MoN$_2$. Using density functional theory calculation, we found multigap superconductivity at temperature up to 36K within anisotropic Eliashberg equation in passivated electride-like surfaces. We also demonstrate their possibility to sustain topological superconductivity with strong spin-orbital coupling.
\end{abstract}


\maketitle

\section{I. INTRODUCTION}
2D superconductors\cite{RN1007} are of interest for their fundamental theoretical particularity, high tunability and potential applications. Their interplay with magnetism, band topology and interfaces is offering a great chance for exotic phenomena. For examples, 2H-NbSe$_2$ is well-known for  coexistence of superconductivity and charge-density wave\cite{RN282,RN1043}, while monolayer h-NbSe$_2$ has recently been identified as Ising superconductor\cite{RN1042,RN1045}; 2M-WS$_2$, with the highest critical temperature $T_c$=8.8K among intrinsic superconductivity in transition-metal dichalcogenides (TMDs), is found to topological nontrivial\cite{RN1046,RN1047}; evidence of topological superconductivity is also observed in Van der Waals heterostructure of NbSe$_2$ and ferromagnetic monolayer CrBr$_3$\cite{RN1048}; the emerging of high $T_c$ in monolayer FeSe on SrTiO$_3$ substrate\cite{RN335,RN38} is intriguing discussion in fundamental mechanism superconductivity along with correlation effect\cite{RN22} and potential antiferromagnetic fluctuation\cite{RN1050,RN1051,RN1052}.\par
In searching for high $T_c$ within Bardeen-Cooper-Schrieffer theory\cite{RN163}, low-mass atoms are indispensable to achieve high Debye frequency and therefore a high $\langle \omega \rangle_{log}$, for examples boron atoms in MgB$_2$\cite{RN941,RN942} and oxygen ions in Ba$_x$K$_{1-x}$BiO$_3$\cite{RN893}. While sufficiently large electron-phonon coupling (EPC) is usually in company with covelent bonding or ligand bonding, of which the bonding energy are sensitive to atomic geometry. In TMD superconductors, polar covalent bonding is expected referring to the difference of electronegativity, but their anions are heavier. A direct extension to light elements should be intrinsic hole-doping Li$_x$NbO$_2$\cite{RN973}, whose $T_c$ is around 10K, higher than all TMDs. Turning to less electronegative anion element, $T_c$ are further promoted to up to 20K in niobium nitrides\cite{RN1054}, intercalated ZrNCl\cite{RN1055} and HfNCl\cite{RN1005}. The recent theortical studies have demonstrate monolayer W$_2$N$_3$\cite{RN1026,RN985} to be a candidate for high-temperature superconductivity with $T_c$ approaching McMillan limit\cite{RN347}. These examples, together with TMD superconductors, are indicating significance of relatively smaller difference in electronegativitiy between cations and anions, or equivalently for generating superconductivity.\par
In this paper, we will demonstrate the possibility of generating superconducting from other experimentally available two-dimensional van der Waals transition metal nitrides (TMNs). Our starting points are MoN$_2$\cite{RN1022,RN1066}, which characterizes in high-oxidation transition metal element. It is hopeful for a good tolerance of electron doping as its capability of ion-absorption have been shown theoretically\cite{RN1016}, so we expect various opportunities for surface modification. Upon this monolayer, transition-metal coating and further passivation are investigated theoretically in aspects of structural stability, electronic properties and EPC strength. Isoelectronic structures of W$_2$N$_3$\cite{RN1026,RN985} are found to have similar superconducting properties with previous theoretical reports. Nontrivial band topology has also be detected and topological edge states are calculated in presence of spin-orbital coupling (SOC) in these systems. With tunable magnetism and superconductivity, and topological edge states close to Fermi level, our findings may provide wide opportunities for realization of exotic electronic states.
\section{II. METHODS \label{methods}}
Quantum Espresso\cite{RN185,RN888,RN889} first-principle calculation package was used to evaluate the structure and properties of proposed materials. In the density functional theory calculation, we chose norm-conserving PBEsol pseudopotential\cite{RN1036} for a better evaluation of short-range correlation effect\cite{RN1037,RN1038}, which shall be important in a d-band metal. Slab model was constructed using z-direction lattice constant being 25$\textup{~\AA}$, leaving a vacuum gap of more than 15$\textup{~\AA}$. For structure optimization and ground state calculation, we sampled Brillouin zone with 12$\times$12$\times$1 uniform $k$ grid. Electronic occupations were estimated using M-V smearing with $\sigma$=0.01 Ry. Energy cutoff for wavefunctions and charge density were set to 100Ry and 600Ry respectively. Dynamical matrix and electron-phonon coupling strength were calculated on 6$\times$6$\times$1 $q$ grid. We further analysised first-principle results with maximally localized Wannier functions\cite{RN1060} to investigate the band topology, and calculated edge states using WannierTools\cite{RN1061}. \par
Experimentally, layered MoN$_2$ has been synthesised through high-pressure high-temperature reaction\cite{RN1022} and is stable under ambient pressure\cite{RN1066}. The unit cell of MoN$_2$ has three rhombohedral MoN$_2$ layers separated by Van der Waals gaps, possessing $R\bar{3}m$ symmetry, with lattice parameters a=2.854$\textup{~\AA}$ and c=15.938$\textup{~\AA}$\cite{RN1022}. Considering vdW being a weak interaction compared to possible direct bonding between N and surface coating metal atoms, we ignored substrates and conducted investigation in a freestanding slab model with a single layer of MoN$_2$. 
\subsection{III. STRUCTURE AND STABILITY}
\begin{figure}
 \includegraphics[width=1.0\linewidth]{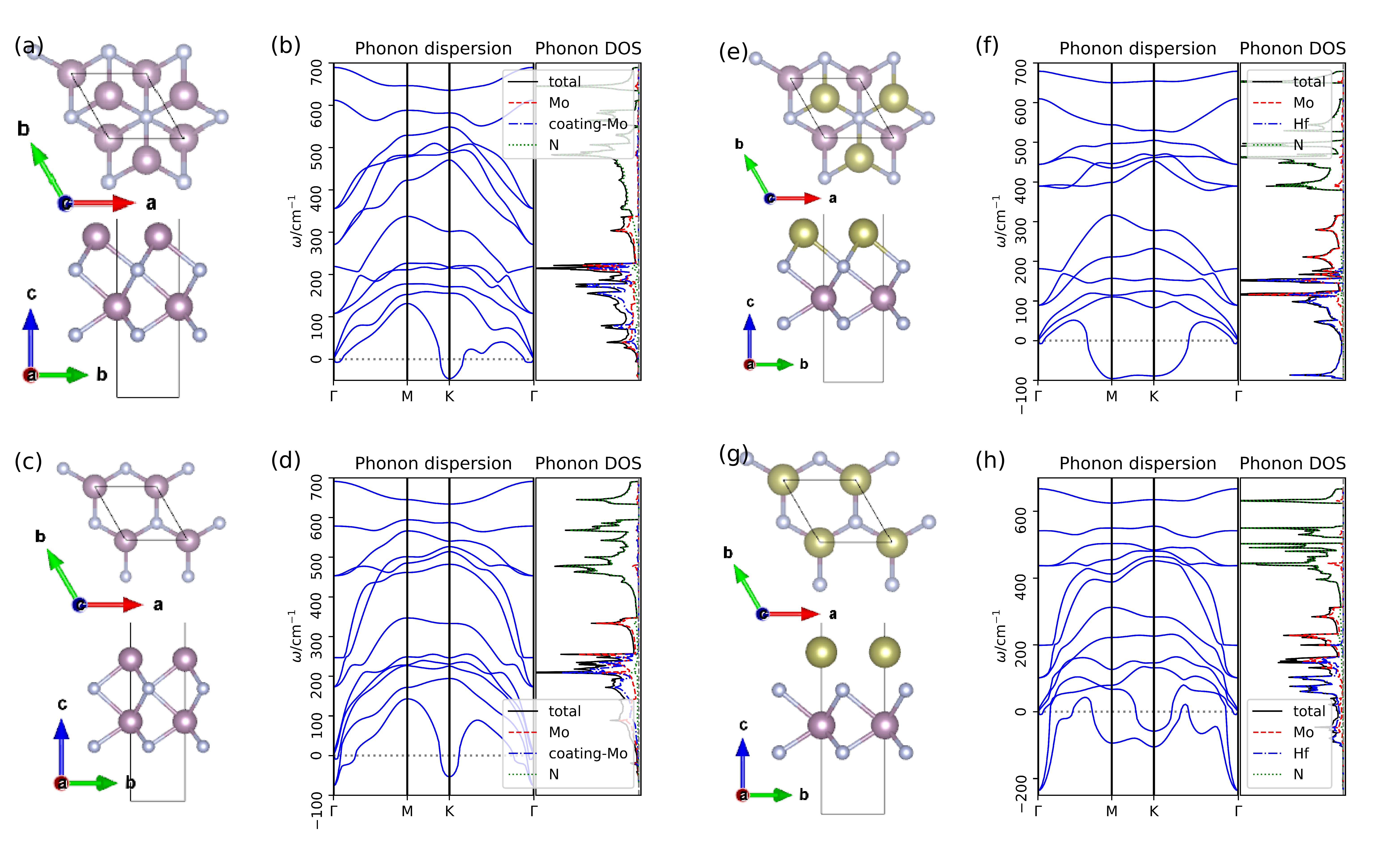}%
 \caption{Structure and phonon dispersion of Mo- and Hf-coated MoN$_2$. (a) and (b) Mo in $o$ site; (c) and (d) Mo in $t$ site;  (e) and (f) Hf in $o$ site; (g) and (h) Hf in $t$ site. We name the structure by the local geometry around the subsurface atom: '$o$' stands for an octahedron around top atom formed by the underlaying atom and the coating atom, and '$t$' for a triangular prism structure. For both coating elements, $o$-sited structure has a lower total energy than $t$-sited one.}
\label{hf-mon2}
 \end{figure}

We began with a test of stability of transition-metal-coated MoN$_2$ monolayer. Assuming one additional transition metal atom per unit cell, two high-symmetry sites  are possible, and we label them as '$o$' and '$t$' here: '$o$' when the additional coating atom forms an octahedron together with underlaying Mo atom around sandwiched  top N atom, and '$t$' when a triangular prism is formed. Our results showed that in general, $o$-site structure has a lower total energy than $t$-site one. The energy difference between two kinds of structure  is 33meV for Hf, 385meV for Mo, 238meV for Nb and 155meV for Ta. For VB group elements, we found $t$-site coating is not only favorable in energy but also processes dynamical stability. Especially, phonon spectrum of Nb-coated monolayer shown phonon-softening at particular $q$, indicating Nb-MoN$_2$ is likely to develop charge-density wave phase. While for Hf and Mo, the structure with lower total energy is found to be dynamically unstable, as shown in Fig.\ref{hf-mon2}. \par
\begin{figure}
 \includegraphics[width=0.5\linewidth]{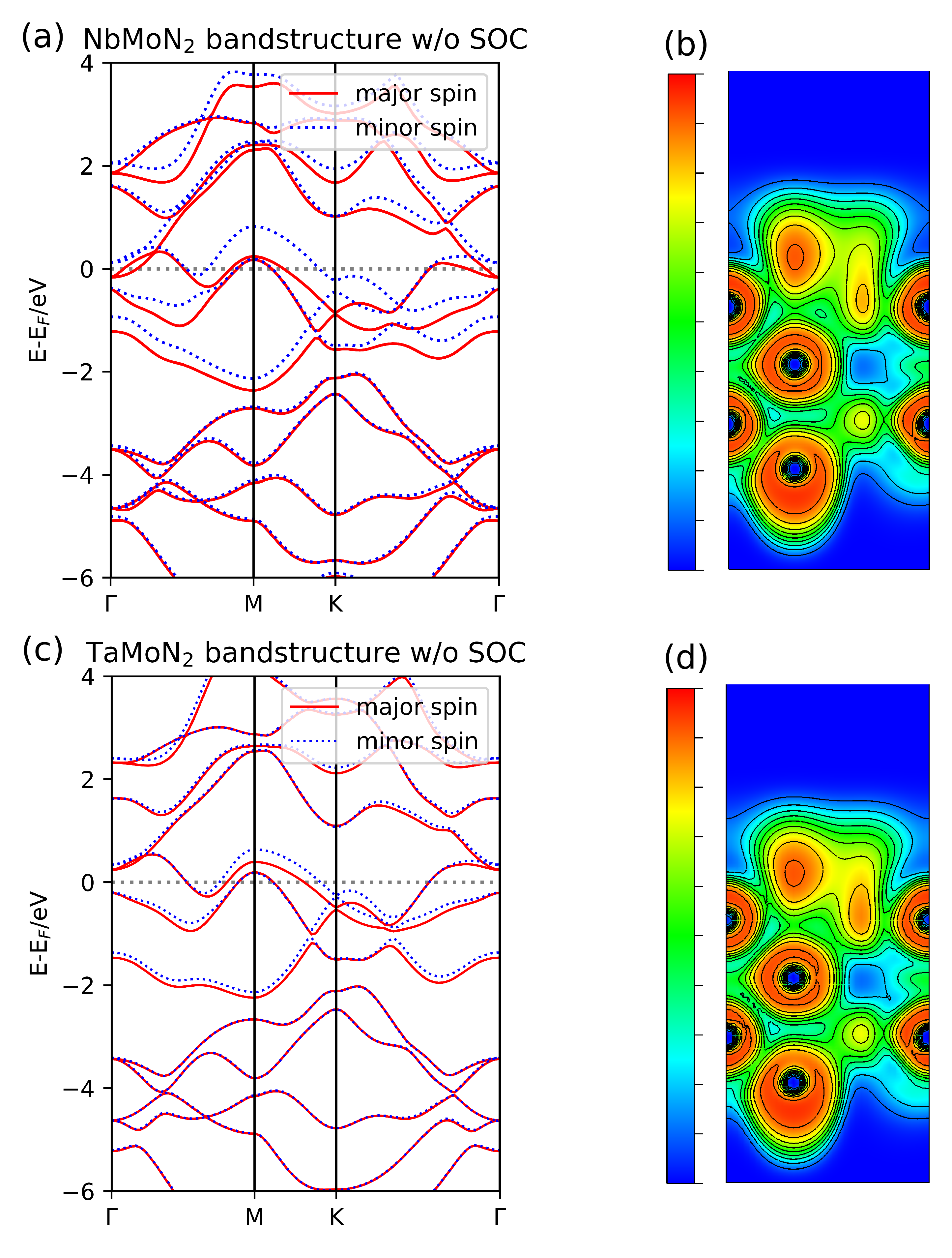}%
 \caption{(a) Spin-polarized band structure of ferromagnetic NbMoN$_2$. (b) Cross section of electron localization function (ELF) of NbMoN$_2$, contour is plotted with an interval of 0.1. (c) Spin-polarized band structure of ferromagnetic TaMoN$_2$. (d) Cross section of ELF of TaMoN$_2$.}
\label{elf}
 \end{figure}

The relaxed lattice parameter for Nb-MoN$_2$ and Ta-MoN$_2$ are 2.854$\textup{~\AA}$ and 2.854$\textup{~\AA}$, respectively. Compared to 2D layered electride materials, such as Ba$_2$N\cite{RN175} and Ca$_2$N\cite{RN193,RN179}, our proposed stable structures resemble in uncovered metallic element surface and unbonded electrons. We therefore examined the behavior of surface electrons by calculating the electron localization function. As one can see in Fig.\ref{elf}(b) and (d), electrons are observed to localized at top $t$ sites upon coating Nb or Ta atoms, away from existing atoms or bonds. \par
For some other potential layered electride materials, it has been proposed that hydrogenating of surface states would enhance superconductivity\cite{RN1028,RN923}. Following discussion on bare NbMoN$_2$ and TaMoN$_2$, we tested two optional sites for the top oxygen atom to place upon $t$-site coated MoN$_2$.  Again we name two difference configures with the geometry around coating VB group atoms formed  by N and O atoms. In both cases, $t$-site oxygen is lower in energy with a difference of 395meV in NbMoN$_2$O and 481meV in TaMoN$_2$O.  This selectivity of top oxygen position is in coincidence with the position of pseudoatom on bare surfaces, and therefore is likely a direct effect of $d$-orbital hybridization. Hence we expect good thermal stability for both systems. Our relaxed lattice parameter of NbMoN$_2$O and TaMoN$_2$O are 2.911$\textup{~\AA}$ and 2.908$\textup{~\AA}$, a small expansion of 2\% and 1.9\% compared with bulk MoN$_2$. Their thickness are 5.123$\textup{~\AA}$ and 5.135$\textup{~\AA}$, respectively.
\section{IV. MAGNETISM}
Notice that transition-metal electride with extra $d$-orbital electrons, namely Hf$_2$S\cite{RN1000}, is magnetic with in-plane ferromagnetic interaction, similar magnetic order may inhabit in the highly-localized electrons in our surface electride, too. Considering overall correlation effects among transition metals, we adopted that on-site Hubbard term $U$=3.0eV \cite{RN1067} in collinear spin-polarized calculation. We found total magnetic momentum to be 0.44$\mu_N$ and 0.19$\mu_B$ for Nb- and Ta-coated MoN$_2$, respectively. Such a small momentum should not be the result of ionic magnetization. On the other hand, magnetization diminishes when $U$ is set to 0eV for coating atoms, while remains almost unchanged with alternating $U$ for Mo in the MoN$_2$ monolayer. Therefore the magnetization should be produced by the Stoner splitting of metallic surface states. In NbMoN$_2$O and TaMoN$_2$O, as the surface states passivated by oxygen, magnetization are found to have damped, even in presence of on-site Hubbard interaction.\par
\section{V. SUPERCONDUCTIVITY}
\begin{figure}
 \includegraphics[width=1.0\linewidth]{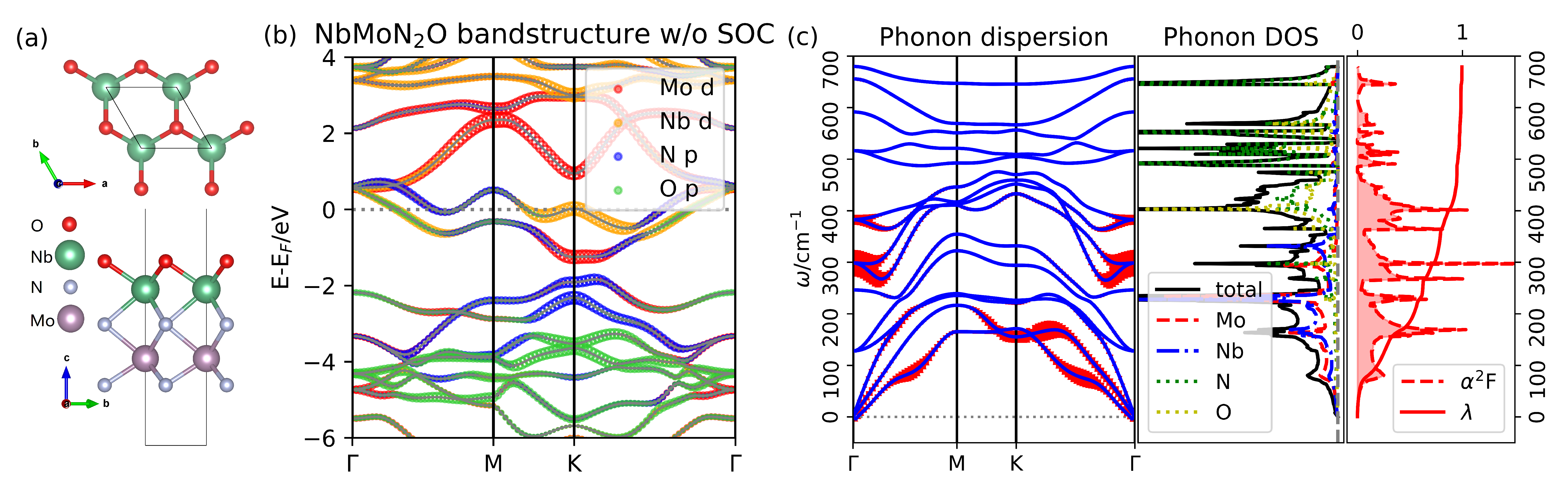}%
 \caption{(a) Structure of oxygen-passivated Nb-MoN$_2$, that is, NbMo$_2$O. (b) Band structure. (c) Phonon dispersion, projected phDOS and EPC $\alpha^2 F(\omega)$ function, mode-decomposed EPC strength are indicated in form of errorbar.}
\label{iso}
 \end{figure}

Given that the surface-passivated NbMoN$_2$O and TaMoN$_2$O are isoelectric to W$_2$N$_3$, which has been predicted to be superconducting, and that our calculation indicates absence of ferromagnetism, we are therefore interested in the potential BCS superconductivity in these systems. The band structure of freestanding NbMoN$_2$O is shown in Fig.\ref{iso}(b). As isoelectronic species of W$_2$N$_3$, its electronic dispersion also resembles that of the later. However, in our case, the bands crossing the Fermi level are of distinct nature. One is original, more dispersive MoN$_2$'s N-Mo hybridized bands, and the another is relatively narrower Nb-O passivated-surface-state bands. Strong hybridization between these two sets of states opens gap at their crossing points.\par 
The dynamical stability of NbMoN$_2$O has been verified through calculation of phonon spectrum, as shown in Fig.\ref{iso}(c). The frequency of acoustic brenches are generally higher than those in W$_2$N$_3$, and no imaginary frequency was detected along high-symmetry paths even with a relatively small smearing (0.01Ry). The EPC parameter $\lambda$ and $T_c$ in NbMoN$_2$O are 1.00 and 20.1-11.0K using McMillan-Allen-Dynes formula\cite{RN1062} assuming $\mu^*$ = 0.10-0.20, comparable to those for W$_2$N$_3$\cite{RN1026,RN985}.\par
\begin{figure}
 \includegraphics[width=0.5\linewidth]{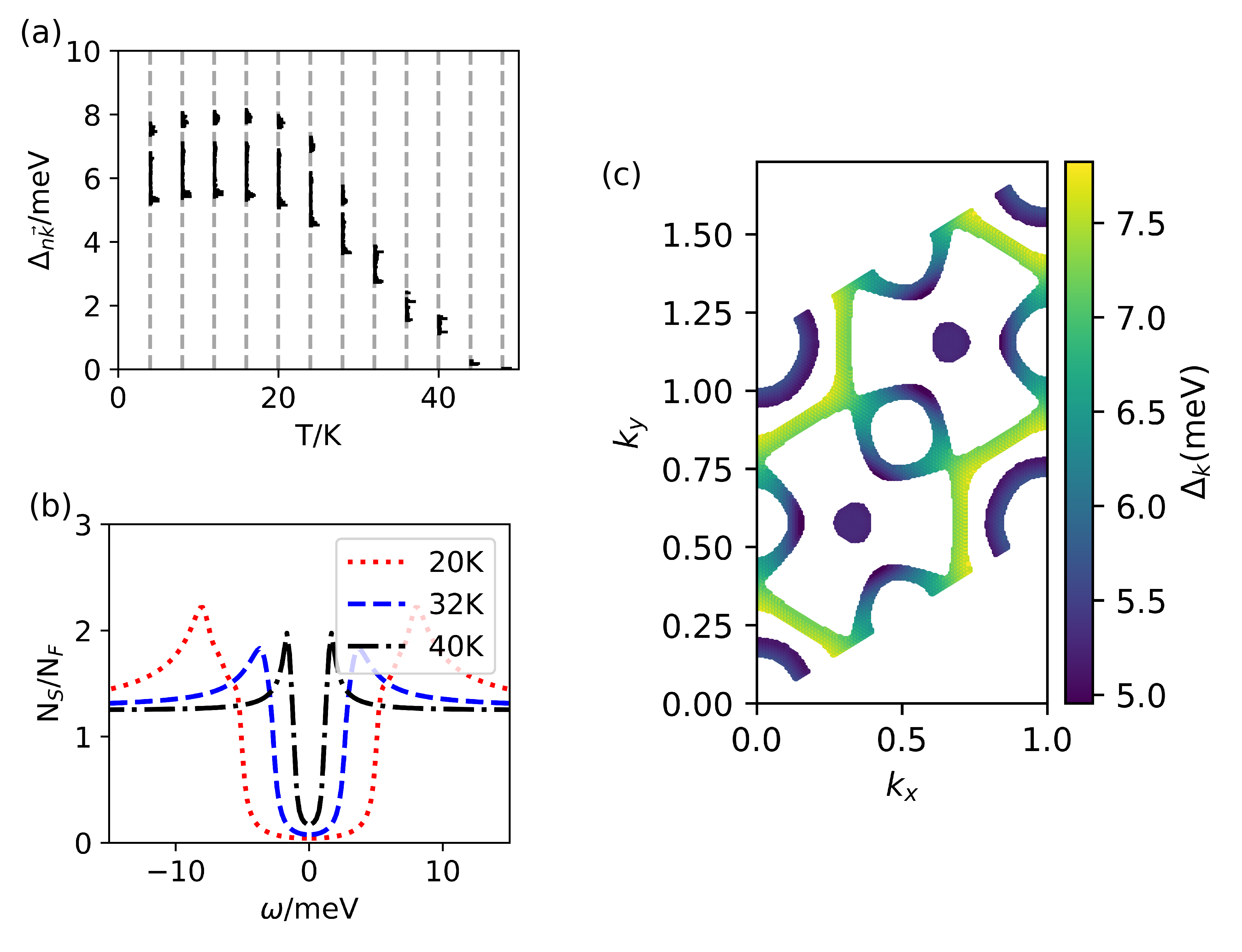}%
 \caption{(a) Temperature-depending anisotropic superconducting gap. (b) Superconducting quasiparticle density of states  $N_S$ normalized by normal states DOS at Fermi level $N_F$ . (c) Momentum-resolved superconducting gap $\Delta_k$ within 0.1eV away from $E_F$, calculated assuming $T$=8K.}
\label{aniso}
 \end{figure}

Multiple Fermi surfaces of distinct nature make it necessary to consider the superconductivity to be of multigap feature, isotropic theory is known to be inaccurate in such a case. Using interpolated dynamical matrix onto 192$\times$192$\times$1 k-mesh and 96$\times$96$\times$1 q-mesh, temperature-depending superconducting quasiparticle gap are calculated by solving anisotropic Eliashberg equation with help of EPW\cite{RN1077}. SOC is omitted here, since the splitting does few modification to the Fermi surfaces expect for hole pocket centering around $K$ point in NbMoN$_2$O, which we will show to contribute little to EPC. The superconducting gap does not vanish up to $T \ge$36K in both systems. Meanwhile, the energy distribution of gap demonstrate multigap nature in this system. In Fig.\ref{aniso}(c) we plot the momentum-resolved gap function $\Delta(k)$ around Fermi level for NbMoN$_2$O at $T$=8K. The largest gap is observed on MoN$_2$-dominating hole-like pocket centering at $\Gamma$, with a size of $\sim$8meV. While for Nb-O-dominating Fermi surface, the superconducting gap is $\sim$5meV. The deviation of isotropic McMillian-Allen-Dynes formula from anisotropic theory may be blamed for unreasonably average of $\lambda$ over all Fermi surfaces. The difference of EPC strength between these two sets of Fermi surfaces is related to their different bonding properties: for Nb-O bond, the difference of electronegativity is quite large and the bonding is ionic, while for Mo-N bond the bonding is polarized covalent, which is sensitive to both bond length and angles. Such a difference is reflected in the band-decomposed charge density of these states. The Mo-N bond has a large weight of charge distribution along the bond than Nb-O.\par
\section{VI. TOPOLOGICAL PROPERTIES}
\begin{figure}
 \includegraphics[width=0.5\linewidth]{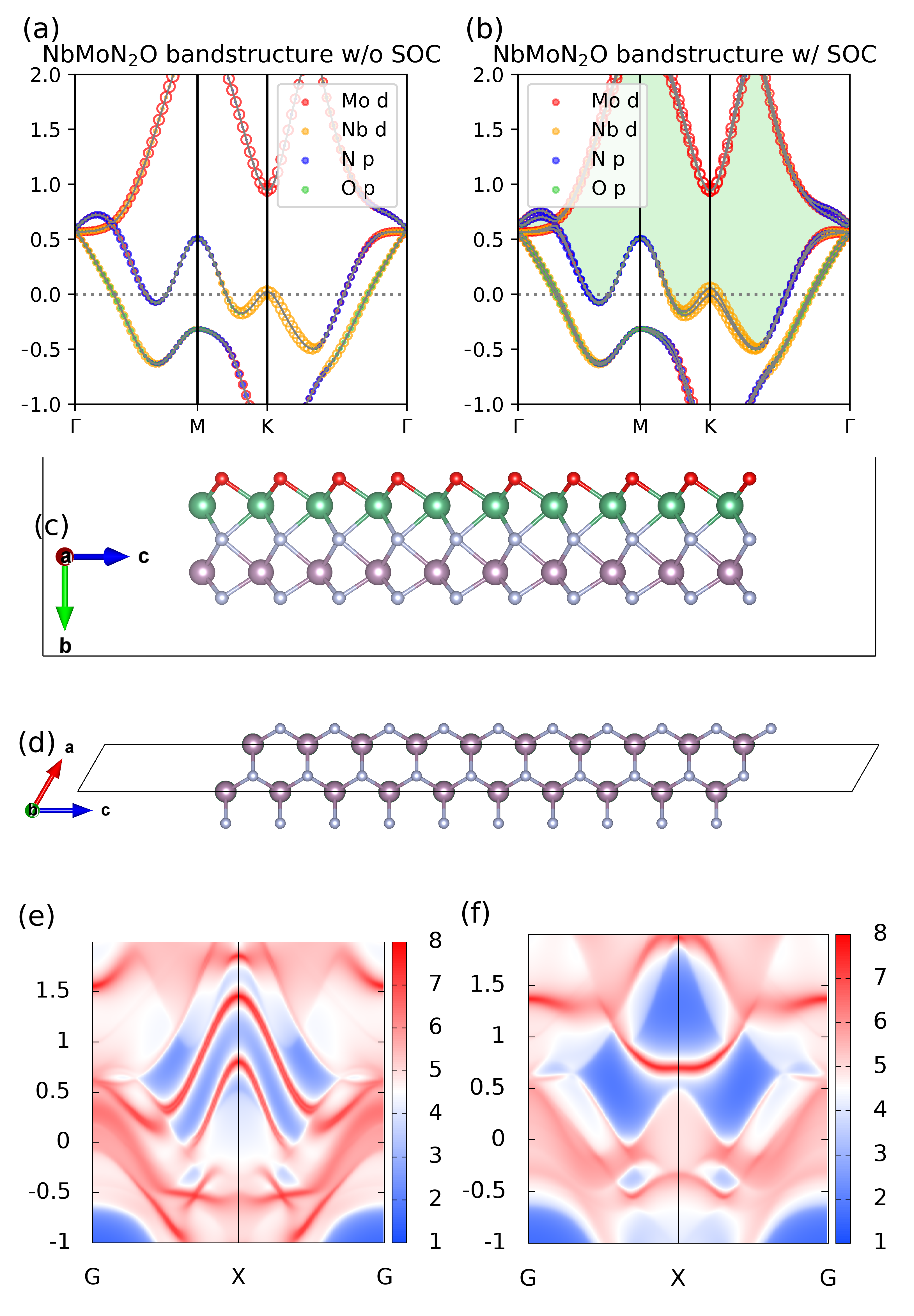}%
 \caption{(a) and (b) band structure of NbMoN$_2$O without and with SOC, respectively. (c) and (d) Momentum-resolved edge density of states on zigzag boundary of NbMoN$_2$O. (e) Slab model with width of 10 unitcell used for edge state simulation, (c) and (d) correspond to the left and right boundary shown in the picture, respectively. Energy in eV.}
\label{topo}
 \end{figure}

Now we explore the topological properties of superconducting surface system NbMoN$_2$O. In the band structure, the degenerate point on the $\Gamma-K$ high-symmetry path seen in scalar relativistic calculation is removed by SOC, leading to an almost global gap settling about 0.6eV above Fermi level. This leads to the inversion between $d$-orbital band and $p$-orbital band and produces nontrivial topology therefore. To analysis the band topology, Wannier function projection and disentanglement was performed starting from all 10 $d$-orbitals of transition metal cations and 9 $p$-orbitals of anions. As shown in Fig.\ref{topo}, using a 10-unitcell-wide slab model, edge modes along the zigzag boundary are simulated. The edge states belonging to the cation-terminating edge, though do not cross, but very close to the Fermi level. We expect that certain electron doping, for example, oxygen-vacancies on the surface, may help to lift Fermi level and enable superconducting topological edge modes.
\section{VII. Discussion \label{discussion}}
The absence of ferromagnetism in the passivated-surface systems allowing us to calculate superconductivity is not only of technical importance, but also is impliciting an inherent connection to superconductivity. To achieve strong EPC, the states constituting the Fermi surface should be sensitive to deformation of atomic structure, that is, atomic orbital to be strongly affected by neighboring atoms. States of covalent bonding is preferred as their wavefunctions have sizable weight around neighbors' valence orbitals, where the strongest atomic field is available for bonding electrons. In contrast, ionic bonding leaves electrons in perturbated atomic orbitals, much less sensitive to neighbors' position. Covalent metals, such as MgB$_2$\cite{RN941} and heavily-doped carbon\cite{RN915}, are good superconductors, with their EPC mainly produced by $\sigma$ bond states. Therefore, it is easy to understand why metallic transition metal nitrides\cite{RN1054} should be superconducting as well as have high hardness, if one refer to electronegativity and conclude their bonding being polarized covalent. Also, covalent bonding requires a pair of spin-antiparallel electrons, which is not likely to support ferromagnetism. Though we have not investigated here, actually, antiferromagnetic interaction may be present. This possibility should not lead to magnetic order due to geometric frustration, but may affect superconductivity in sense of other bosonic excitation, such as spin fluctuation in analogy to FeAs-based materials\cite{RN368,RN1075}.\par
Technical difficulties are there when dealing with out-of-plane acoustic mode. In a freestanding 2D model, retracting forces of out-of-plane acoustic mode are mainly determined by changes in in-plane bond length, as in z-direction atoms are moving synchronously. This results in the retracting forces being parabolically depending on out-of-plane displacement and softer than in-plane acoustic modes. Such an instability is well-known in 2D systems, but does not actually affect the stabilization of real systems where substrates play an important role. Moreover, we have observed the imaginary frequence area shrink when density of both k-mesh and q-mesh are increased. We therefore conclude these imaginary-frequency modes are numerical errors limited by current techniques and not harmful to stability.\par
In conclusion, within framework of density functional theory and anisotropic Eliashberg equation, we have suggested a strategy of realizing both nontrivial topology and high-temperature superconductivity in a 2D system, from an experimental available layered transition metal nitride MoN$_2$. VB group elements, namely Nb and Ta are able to form stable monolayer coating upon hexagonal MoN$_2$, and electride-like surface state and ferromagnetism are expected to be present. The nontrivial topology arises from the strong SOC provided by the transition metal elements. By passivating the surface states, ferromagnetism is removed and we found $T_c$ up to 36K within anisotropic Migdal-Eliashberg theory. Together with early studies on W$_2$N$_3$, our results are indicating TMNs' potential to push to McMillian limit of superconductivity.
\nocite{*}
\bibliography{ref}

\newpage

 \begin{figure}
 \includegraphics[width=1.0\linewidth]{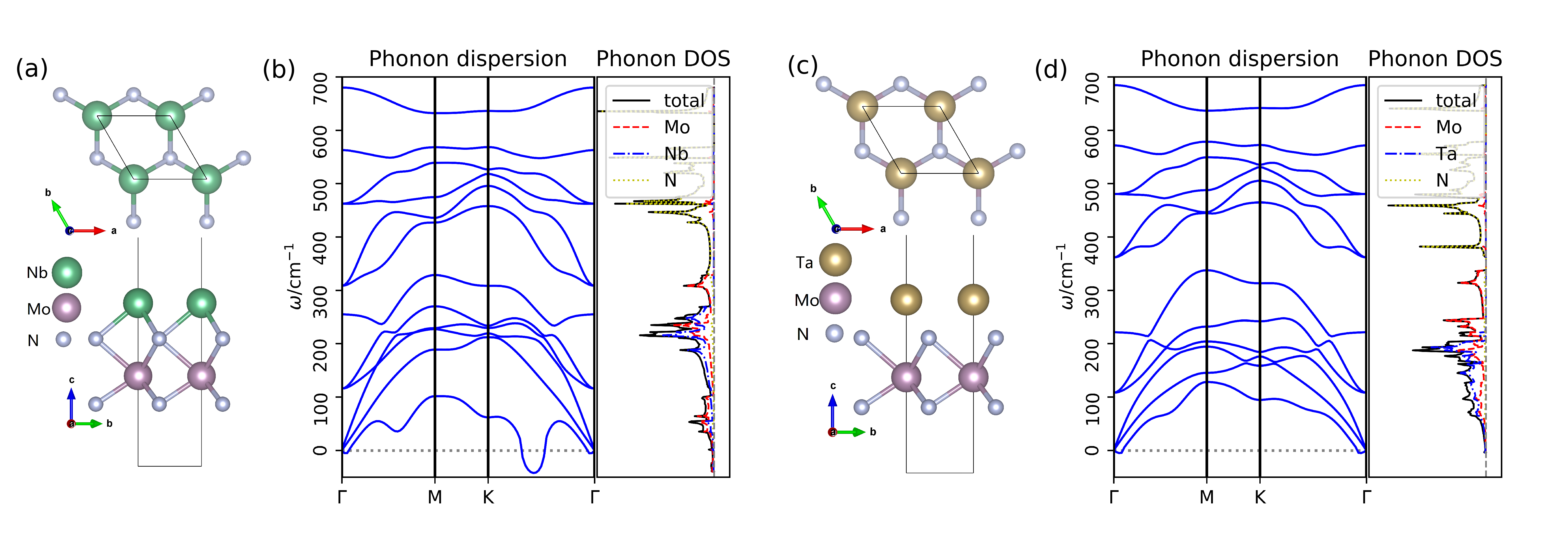}%
 \caption{(a) Atomic structure of $t$-site Nb-coating MoN$_2$. (b) Phonon dispersion and phononic density of states (phDOS) of NbMoN$_2$. (c) Atomic structure of $t$-site Ta-coating MoN$_2$. (d) Phonon dispersion  and phDOS of TaMoN$_2$.}
\label{bare-ph}
 \end{figure}

\begin{figure}
 \includegraphics[width=0.5\linewidth]{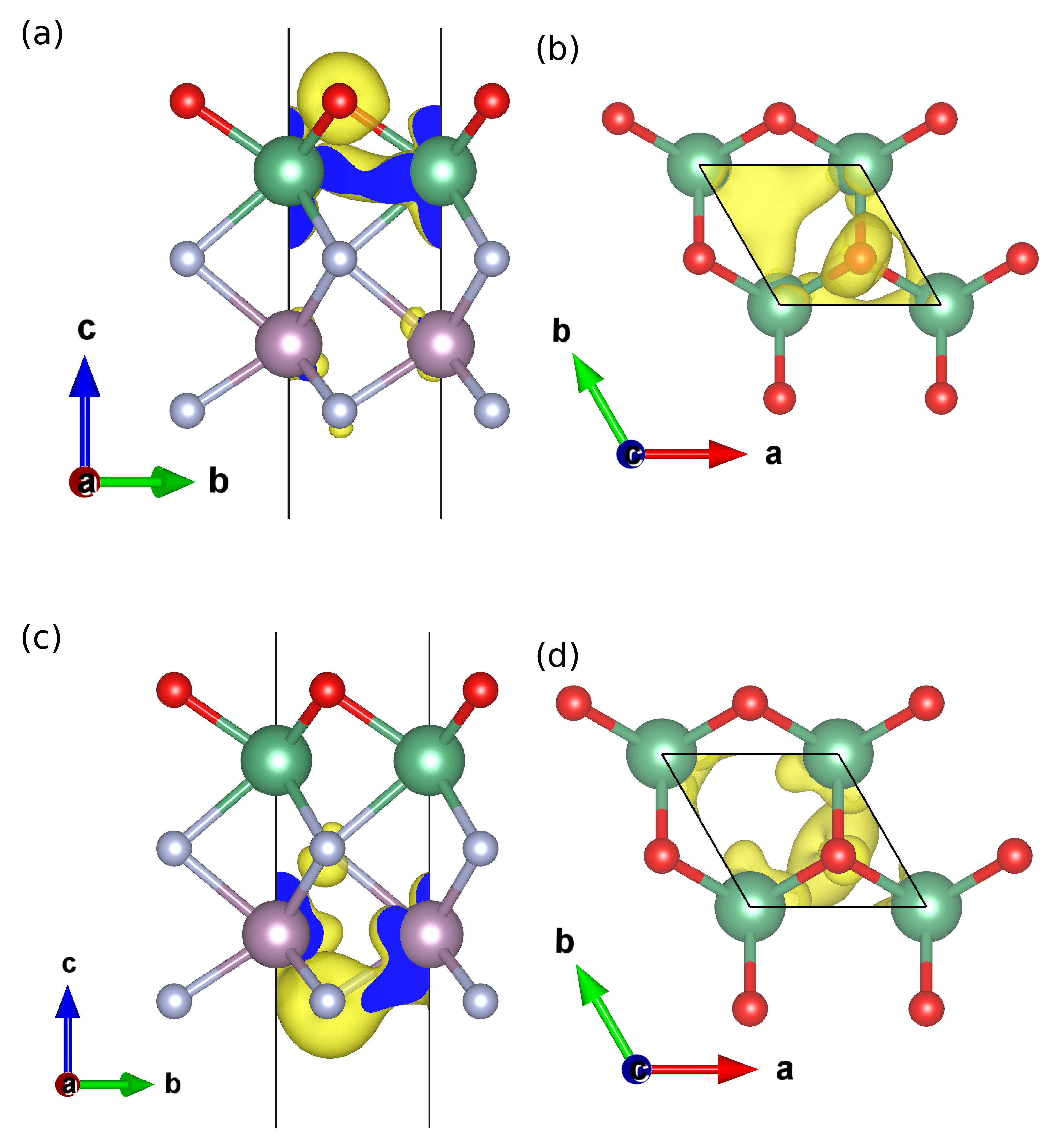}%
 \caption{Band-decomposed charge density of Nb-O bond state (a) side view; (b) top view; And of Mo-N bond state (c) side view; (d) top view. Note charge density damps to zero around the bond center for Nb-O bonds, but remains finite for Mo-N bonds.}
\label{bond}
\end{figure}

\end{document}